\documentclass[10pt]{article}

\usepackage{amsmath}
\usepackage{amssymb}

\usepackage{graphicx}

\usepackage{cite}

\usepackage{color} 

\usepackage[labelfont=bf,labelsep=period,justification=raggedright]{caption}

\makeatletter
\renewcommand{\@biblabel}[1]{\quad#1.}
\makeatother

\date{}

\begin{document}

% Title must be 150 characters or less
\begin{flushleft}
{\Large
\textbf{The weakness of `weak ties' in the classroom}\\[0.1in]
}
% Insert Author names, affiliations and corresponding author email.

Luis M. Vaquero$^{1}$ and 
Manuel Cebrian$^{2}$\\[0.1in]

\it \small {$^1$} Hewlett-Packard Laboratories, Bristol BS34 8QZ, UK
\\
{$^2$} Department of Computer Science and Engineering, University of California, San Diego, CA 92093, USA
\\
\end{flushleft}

% Please keep the abstract between 250 and 300 words
\subsection*{Abstract}

Granovetter's ``strength of weak ties'' hypothesizes that isolated social ties offer limited access to external prospects, while heterogeneous social ties diversify one's opportunities. We analyze the most complete record of college student interactions to date (approximately 80,000 interactions by 290 students -- 16 times more interactions with almost 3 times more students than previous studies on educational networks) and compare the social interaction data with the academic scores of the students. Our first finding is that social diversity is negatively correlated with performance. This is explained by our second finding: highly performing students interact in groups of similarly performing peers. This effect is stronger the higher the student performance is. Indeed, low performance students tend to initiate many transient interactions independently of the performance of their target. In other words, low performing students act disassortatively with respect to their social network, whereas high scoring students act assortatively. Our data also reveals that highly performing students establish persistent interactions before mid and low performing ones and that they use more structured and longer cascades of information from which low performing students are excluded. 
% Please keep the Author Summary between 150 and 200 words
% Use first person. PLoS ONE authors please skip this step. 
% Author Summary not valid for PLoS ONE submissions.   
%\section*{Author Summary}

\subsection*{Introduction}

Computer Supported Collaborative Learning (CSCL) requires appropriate methods for evaluating collaboration in a way that researchers and professors can gain more insight into the results of innovative experiences and
lecturing/teaching procedures \cite{Neale99}. However, systematical gathering and analysis of educational data \textit{in-vivo} has only recently started. The literature highlights the key role of student interaction for effective learning. Interaction can take place by using many different tools and frameworks, which have proven to be useful for the evaluation of student's performance \cite{martinez2003combining}. The study of the social and participatory aspects of learning is an ideal arena for social network analysis techniques \cite{Wasserman94}. The analysis of social networks has demonstrated to be a valuable tool for CSCL applications, indeed, achieving desirable learning outcomes requires an appropriate social network \cite{Morrison02}.

Most studies focus their analysis on structural features of the network, such as node centrality. For instance, Nurmela {\it et al.} looked at the structure of the interactions trying to determine the central actors of the CSCL environment \cite{nurmela1999evaluating}. In this social structure, ``key communicators'' were assumed to be the most connected individuals \cite{cho2007social}. Similar analyses were carried out by Mart\'inez {\it et al.} \cite{martinez2003combining} and Chen and Watanabe, who focused on other parameters important for the final score: group structure, member's physical location distribution, and member's social position \cite{Chen07}.

While it seems clear that the relevance of the network structure and interactions has been widely recognized \cite{Sundararajan07}, some other factors (e.g. social acceptance or willingness to communicate) affecting the dynamic interaction patterns of the classroom have recently been recognized as essential ingredients \cite{Yu10}. Granovetter's pioneering work recognized the importance of interaction patterns and proposed his well-known ``strength of weak ties'' phenomenon, where he hypothesized that isolated social ties offer limited access to external prospects, while heterogeneous social ties diversify one's opportunities~\cite{granovetter1973strength}. Recent empirical work confirmed that the diversity of individuals' relationships is indeed strongly correlated with the economic development of communities~\cite{eagle2010network}. 

It is tempting to make a direct extrapolation to educational contexts. However, understanding whether this relationship also holds in the classroom may provide insight to help shape better educational strategies. In general, it is not just about knowing who students interact with, but how and when they do it and, importantly, what is the result of these interactions for the educational outcome~\cite{Ullrich10}.

Preliminary answers to the ``how'' come from several works that analyzed the ``macroscopic'' effects (effects on structure) depending on relationships reconstructed from the messages sent \cite{Yeung05,Kepp09} or also considering the type of interaction being held \cite{Cho07}. This trend on acquiring knowledge from interactions was also followed by Erlin {\it et al.}, who considered the content under discussion in addition to the interactions themselves \cite{Erlin08}. Most of the  previous analyses correspond to a given static snapshot of the network at some point in time or a reduced number of samples, for instance, \cite{martinez2003combining} analyze these macroscopic metrics in the four different assignments the course was structured in ($\simeq$ once a month).

This paper tries to gather details on the dynamics and mechanics of collaboration, by characterizing the type of interactions on temporal terms and relating these types with the final outcome of the course (student score). We also aimed to answer the ``when'' question by characterizing network evolution at a microscopic level (interaction level) at unprecedented temporal resolution. We hypothesize that gaining insight into these data could be a valuable tool to reduce course dropout rates. More than 1.2 million students drop out of school every year in the U.S., one every 26 seconds (per day figure derived by dividing 1.23 million by 180 school days per year. Per second figure derived by dividing 1.23 million by 31,536,000 seconds in a full calendar year \cite{dropout1}). 2007 dropouts will cost more than \$300 billion in lost wages, taxes and productivity to the U.S. Dropouts contribute about \$60,000 less in federal and state income taxes. Each cohort of dropouts costs the U.S. \$192 billion in lost income and taxes \cite{dropout2}. A dropout student is more than 8 times as likely to be in jail or prison as a high school graduate and nearly 20 times as likely as a college graduate\cite{dropout3}. 

The rest of this paper is organized as follows. The next Section presents the main results obtained from our analysis. This exposition is followed by a discussion. The materials and methods employed for data analysis are detailed in the final section of the paper.

% Results and Discussion can be combined.
\subsection*{Results}

We analyzed the most complete record of college student interactions\footnote{See Course Details in the Materials and Methods scetion for a concrete definition on what we mean by interaction} to date and compared the social interaction data with the academic scores of the students. To this end, we analyzed records of $~80,000$ interactions by $290$ students --- approximately $16$ times more interactions with almost $3$ times more students than previous studies on educational networks. The data cover a high resolution of both social interactions  in the classroom and out of the classroom (see Materials and Methods for more details), being independent of gender differences (correlation of gender to score was -0.04). Figure \ref{fig:data}A shows the social graph for one of the classes being analyzed. 

\subsubsection*{Diversity and Assortativity Analysis}

Our first finding is that social diversity is negatively correlated with performance. This is explained by our second finding: highly performing students interact in groups of similarly performing peers. This effect is stronger the higher the student performance is. Indeed, low performance students tend to initiate many transient interactions no matter the performance of the students they interact with. In other words, low performing students act disassortatively with respect to their social network, whereas high scoring students act assortatively. In the following we give details of these findings. 

We start by comparing the score of each student with diversity metrics associated with the interactions held by each member of the social network (as shown in Materials and Methods). 

% We characterize the nature and diversity of interaction ties within an individual's social network. Specifically, topological diversity is caculated as a function of Shannon's entropy. Then, social diversity is defined as Shannon's entropy associated with individual communication behavior, normalized by the total number of interactions 

\begin{figure}
\caption{A: shows a graph of one of the analyzed courses including 82 students. B: Scatter plot and linear regression for one of the variables analyzed (number of interactions) vs. scoring in one of the classes ($R^{2}=0.72$). C: Scatter plot and linear regression for social diversity vs. scoring in one of the classes ($R^{2}=0.12$). D: Normalized social diversity when students were grouped according to their performance. $^{*}  p<0.05$ as compared to low;$^{\#} p<0.05$ as compared to mid; n=290.}
\begin{center}
\includegraphics[width=0.85\textwidth]{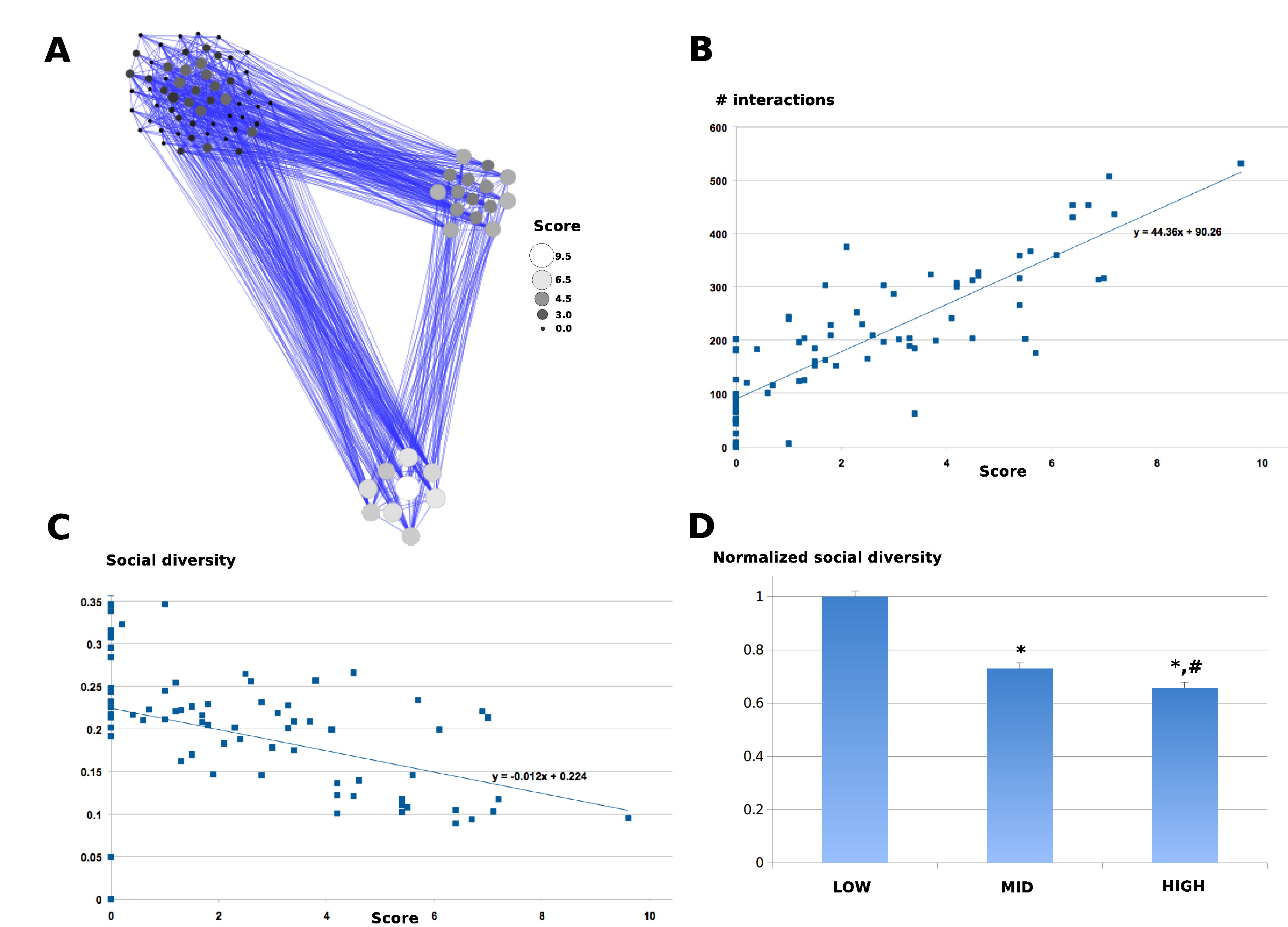}
\end{center}
\label{fig:data}
\end{figure}

The number of connections (students a student has interacted with), number of interactions (times a user has contacted or been contacted with/by other student) and the topological diversity (a function of Shannon's entropy, see Materials and Methods) were all positively correlated with the final score of the student (Pearson's correlations of $0.81, 0.85, 0.74$, respectively; $p<0.01$), as shown in Figure \ref{fig:data}B. Principal component analysis of these metrics revealed that all of them were closely interrelated, resulting in non-significant improvement when combined (see Materials and Methods). However, social diversity negatively correlated with final scores ($-0.34, p<0.01$) (Figure \ref{fig:data}C), a more diverse number of interactions resulting in a reduced score. 

To further analyze the effects on score, students were grouped into high ($>6.5$), mid (between 6.5 and 3.5) and low ($<3.5$) scoring. To verify the suggested existence of less effective interactions (Figure \ref{fig:data}C), we also classified the type of interactions in two types: 1) persistent, those sustained in time, and 2) transient, those never repeated. We find that $38 \pm 12\%$ of the interactions held by highly performing students were persistent, which is statistically different to those held by mid ($17 \pm 5\%$) or low ($2 \pm 2\%$) performance students ($n=290, p<0.05$). 

We analyzed the average number of persistent interactions per neighbor: a higher number indicating more targeted interaction to a reduced number of neighbors. This is illustrated in Figure \ref{fig:data2} in the Materials and Methods (top panel) for one of the three classes under analysis. 

The presence of more focused and sustained interactions did not preclude high scoring students from interacting with colleagues with mid or low scores in a transient manner (similar number of transient interactions regardless of the score, see Figure \ref{fig:data2}, bottom panel). An assortativity analysis \cite{newman2003mixing} ($r=0.5, p<0.05$ by using the Jackknife method, see Materials and Methods) on these persistent interactions indicated the existence of preferential interaction initiation. In other words, similarly scoring students tended to keep persistent interaction between themselves.

\subsubsection*{Temporal Analysis}

\begin{figure}[htp]
\centering
\caption{Temporal evolution of the total number of interactions in all groups. The ``y'' axis indicates the number of interactions per group per week normalized to the value of the weeke when the maximum number of interactions was recorded for that group.}
\includegraphics[width=0.6\textwidth]{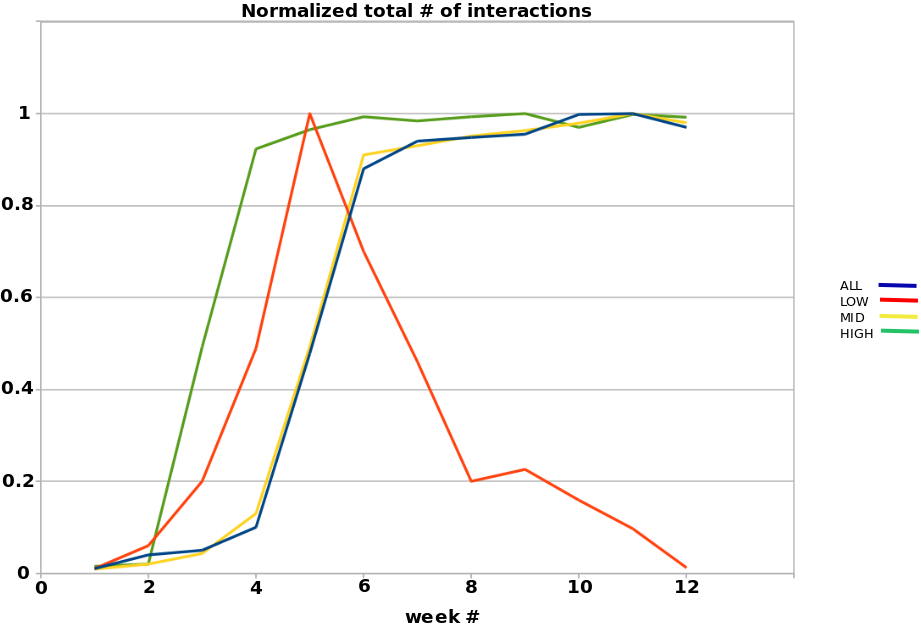}
\label{fig:temporal}
\end{figure}

One interesting finding is that the total number of interactions per week (normalized to the maximum value in all weeks) for all groups increases over time and it saturates around week $6$ for mid and high performing students and around week $4$ for high performing student (Figure \ref{fig:temporal}). In both cases, the number of persistent and transient interactions increase as the week number increases until saturation. However, the number of interactions for low scoring students behaves in a strikingly different manner. The number of total interaction increases until week 4, where it starts dropping steadily until the end of the course (Figure \ref{fig:temporal}).

\begin{figure}[htp]
\centering
\caption{Evolution of the \% of persistent interactions (relative to the average total \# of interactions of that group) per week and per student group (low, A; mid, B; and high, C) relative to the total number of interactions per group per week. Continuous lines represent the fit of a curve to the points as indicated in Section on Materials and Methods.}
\includegraphics[width=0.6\textwidth]{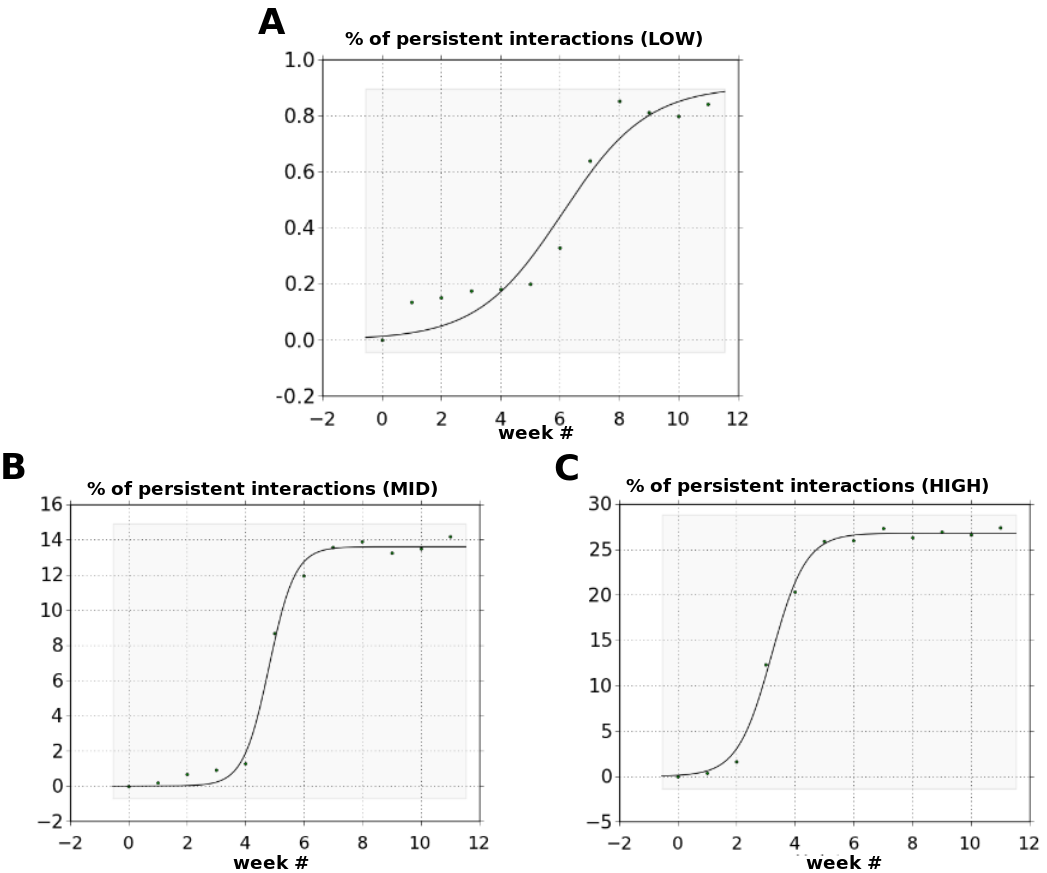}
\label{fig:temporal2}
\end{figure}

A closer look at the data reveals that the percentage of persistent interactions increases in all groups, but at different rates (\ref{fig:temporal2}A, B, and C).

%This increase was statistically significant for mid and high scoring students (Figure, remaining almost flat for LOW scoring students (less than 1 \%). 

As indicated in the table in Materials and Methods, the midpoint for the sigmoid function was $6.08$, $4.81$ and $3.2$ for low, mid and high performing students ($p < 0.05$). This indicates that high performing students on average establish persistent interactions before mid and low performing students ($1$ and $2$ weeks before, respectively). Also, mid performing students start to establish persistent interactions $1$ week before low performing students do. If one takes the slope of the sigmoid as a reference, it can be observed that there is no significant difference in the rate of change from a ``low interaction mode'' to a ``higher interaction mode'' between mid and high performing students ($0.58$ vs. $0.4769$).

Taking these data on increasing \% of persistent student interactions with the assortivity analysis (students prefer to interact with students in their own group) above is suggesting that at some point reciprocity $ R_{i,j}$ (measured as the fraction of times a student $i$ in any given group responds to a student $j$ outside her same group) may start dropping. Reciprocity remained unchanged with time and was similar between groups ($\simeq 0.7$), suggesting that even when high performing students do not usually initiate interactions with low performing ones, they answer back when they receive some request. 

This could be indicating that low performance is due to a lack of interest of the students or just that no valuable content was conveyed in these ``forced'' interactions. Since the content of these interactions was not logged, we needed to find other mechanism to determine how valuable content flows between students and groups of students.

\subsubsection*{Information Cascades}

%\cite{Leskovec2006,Leskovec2006b}
Information cascades reveal spread mechanisms in which an action or idea becomes widely adopted due to the inﬂuence of others, typically, neighbors in some network like cascades in the context of a large product recommendation networks. In order to detect the presence of information cascades and determine the actual value of the communication, we needed to gain insight on the content of the messages exchanged by students. Since this would be a clear violation of students' privacy, we decided to analyze another source of information: file exchange of students in their home directories and in their BSCW accounts (see ``Information Cascades'' in Material and Methods below). 

\begin{figure}[htp]
\centering
\caption{Most Frequent Cascades for Low Performing (\textbf{A}) and High Performing (\textbf{B}) students. Students initiating, relaying or receiving a document were supposed to be part of the cascade.}
\includegraphics[width=0.7\textwidth]{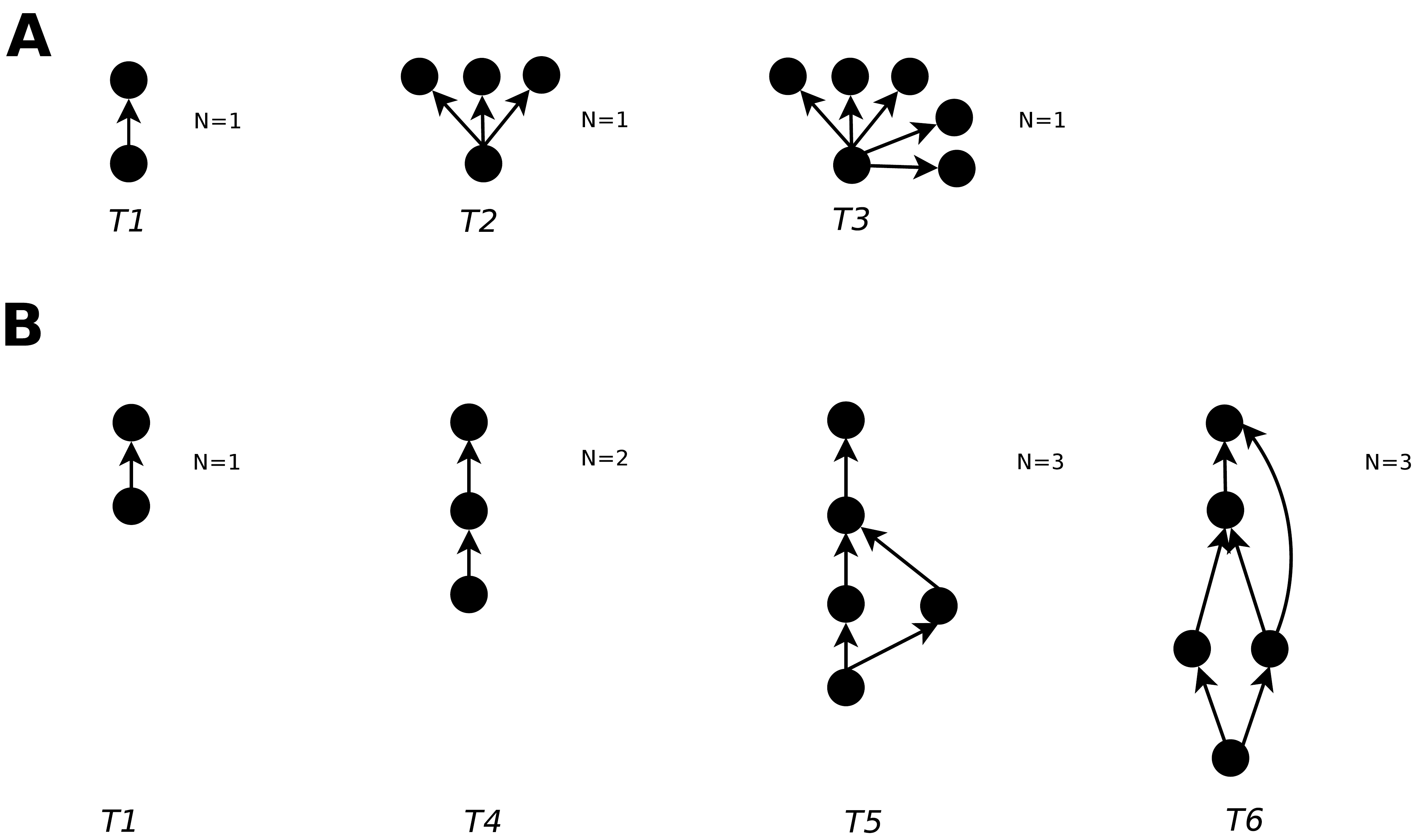}
\label{fig:cascades}
\end{figure}

We define as trivial cascades those implying a single transfer (a single originating source and a single destination) of information about the course, and non-trivial cascades, those with more complex patterns. We found a total of $845$ cascades, and $53.37\%$ of which were trivial cascades (\textit{T1} in Figure \ref{fig:cascades}), $25\%$ are non-trivial cascades involving transfer from a single source to many destinations in the same time frame, and the remaining $11\%$ of the cascades are topologically more complex.

The number of cascades is significantly different across all three groups $51\%$, $35.97\%$ and $13.03\%$ for high, mid and low performing students, respectively (see Table \ref{table:cascades}). 

Our data reveal that the length of the cascade (number of synchronous transfers) gradually increases as the average score of the students involved in the cascade increases. This is also supported by the fact that among non trivial cascades, the most common pattern for low performing students was star-like (\textit{T2 and T3} in Figure \ref{fig:cascades}, $97.8 \%$), while chained cascades (\textit{T4, T5 and T6} in Figure \ref{fig:cascades}) were more common for mid ($53.82\%$) and high ($76.29\%$) performing students.

\begin{center}
\begin{table*}[t!]
\centering
\begin{tabular}{|l|r|r|r|}
\hline
 & \textbf{LOW} & \textbf{MID} & \textbf{HIGH}\\ 
\hline
\textit{\% of Cascades} & 13.03 & 35.97 & 51\\ 
\hline
\textit{\% trivial} & 96.36 & 65.13 & 34.1 \\
\hline
\textit{Average Cascade length} & 1.05 & 1.95 & 2.9\\ 
\hline
\textit{\% stars} & 97.8 & 46.18 & 23.71 \\
\hline
\textit{\% chain} & 2.2 & 53.82 & 76.29\\\hline
\end{tabular}
\caption{Summary of the cascade analysis performed across the three groups of students.}
\label{table:cascades}
\end{table*}
\end{center}

\subsection*{Discussion and Conclusion}
\label{sec:discussion}
Combining data from a large educational network with each student's individual score, we were able to gain insight into the following question: Do more diverse ties imply better academic performance? 

Our results show that a higher interaction number (independently of the number of distinct students involved) is usually an indicator of higher score. However, increased social diversity is negatively correlated with high scores, which indicates that not all the interactions are equally productive. The higher the score of the students, the higher the percentage of their iteractions that were persistent. 

As the score of the student increases, these persistent interactions are initiated with a reduced number of similarly performing colleagues (assortative interaction pattern). Low performing students have a larger number of transient interactions spread over a large number of neighbors. Social network diversity seems to be at the very least a strong structural signature for the (negative) academic performance of students. 

The fact that the number of interactions per week increases as the course progresses may indicate that students gain confidence in the course methodology and tools. The dynamics of these interactions reveal that once students start establishing persistent interactions they do it more and more until a maximum saturation point is reached. Highly performing students tend to initiate persistent interactions before lower performance ones, suggesting a higher willingness to collaborate. A striking fact is that these highly performing students still maintain more than $\simeq 70\%$ of transient interactions, mostly with mid performing students. Our reciprocity analysis shows that students try to contact high performing students and these feel some sort of obligation to respond. 

Evidently, we could not monitor the content of the private message of students and decided to perform an information diffusion analysis that could help us to gain insight on the content being actually exchanged. Our results reveal that low performing students generally exchange documents in a trivial manner (i.e. in a forwarding manner that spans a single hop). On the contrary, more complex and longer cascades occur in highly performing groups. This indicates the existence of a highly organized network where similarly performing students exchange information in a well-structured fashion, following characteristic patterns that are different across groups. While highly performing students mainly exchange documents in a chained manner, low performing students spread the information to many other students at the same time, without this document apparently being relayed to other students beyond the recipient. Indeed, low performing students were not typically included in the information chains developed by high performing students. By this we do not mean to imply a mean behavior by students, but most likely it is indicating the presence of a benefit maximization process by which students focus their efforts in potentially more fruitful connections.

Getting lower value information is just one side of the picture. Low performing students drastically reduce the number of interactions after week $4$, which is indicating a clear lack of motivation that leads them to drop the course and focus on other tasks. The fact that the percentage of persistent interactions does not significantly increase indicates that these students initiate and drop interactions in an inefficient manner. This \textit{per se} does not let us conclude a lack of skills or motivation by low performing students. 

We analysed these data and found that: 1) social diversity is a strong indicator of low performance and it is linked to weak interactions; 2) low performing students are not typically included in the highly-structured information exchanges held by highly performing colleagues; 3) low performing students drastically reduce the number of interactions they held. These three elements may be the causes or effects of a de-motivation phase in low performing students (studies targeted at detecting causality relationships between these three and scoring are needed). 

As part of our future work, we hypothesize that detecting this dropping behavior early in the course and getting low performing students involved in high performing chains could help increase the final score of the students. Endowing educators with tools to allow them to pay additional attention to those more likely to drop their interactions and help them to focus on who they interact with. On the other hand, this may have a negative effect on highly scoring students who will get many more interactions they will feel obliged to respond to. Such a tool could result in huge benefits for the society in terms of reduced exclusion of individuals and also in economig terms \$60,000 less in federal and state income taxes and \$192 billion in lost income and taxes per dropout in the U.S. \cite{dropout2}

%We hope this findings encourage additional research targeted at to deriving reliable causality relationships to determine whether network diversity .

% You may title this section "Methods" or "Models". 
% "Models" is not a valid title for PLoS ONE authors. However, PLoS ONE
% authors may use "Analysis" 

%\newpage

\section*{Materials and Methods}

\paragraph{Course Details}

The data are of the interactions of 290 students at Universidad Rey Juan Carlos, Madrid, Spain, EU during two consecutive years of a 12-week long course on Basic Computer Science Skills (in Linux such as OpenOffice, GIMP, or content licensing techniques such as Creative Commons) for freshmen students of journalism. 

The students were belonged into three groups depending on their year class and on lab room availability. Two groups (79 and 82 students) belonged to the 2010 course and the remaining students were placed in a single group during the 2011 course. Thus, three different graphs were built to ease the analysis process and obtain an average behavior for all the students involved in this study. Students voluntarily signed a collaboration agreement including privacy clauses about their data and specific information on what was going to be kept. 

These data included the logs of class content-related communications between students done via Moodle, a classroom IRC, BSCW and a Canvas Chat application that they included in their Facebook account (up to 35\% of the interactions were done via this canvas). An interaction is defined as a communication attempt via the aforementioned systems. In one-to-many communication mechanisms (e.g. a post in Moodle), the interaction count was increased only if the post received an answer. All students, but one (who was excluded from the study), were frequent Facebook users (daily utilization: $1.5 \pm 0.9 $hours/day; average number of friends in Facebook: $142 \pm 85$). 

We transformed interaction data into a network by defining an undirected edge as an exchange of messages between two nodes, such that each party originated at least one message to the other. We also kept track of the number of interactions that took place over a given connections. The average age of the students was $18.5 \pm 0.8$ years, $65.5 1\%$ were women.

\paragraph{Data Anonymization and University Approval}
Student ids were obscured and randomly re-arranged so that the data analyzer could not track a student. Each recorded interaction was assigned an ID in each one of the employed systems. This timestamp-based ID was replaced by random, unique identifier for the different systems employed in this study. While deductive disclosure is always a possibility with logged interaction data, this provided adequate blinding for the study to acquire university's approval.

\paragraph{Diversity Metrics}  

We used several measures of the diversity in an individuals' social network, including topological diversity, assortativity, and structural holes. We characterize the nature and diversity of interaction ties within an individual's social network. Specifically, topological diversity is calculated as a function of Shannon's entropy ($H(i)=-\sum_{j=1}^{k}p_{ij \log(p_{ij})}$), where \textit{k} is the number of \textit{i}’s contacts and $p_{ij}$ is the proportion of \textit{i}’s total interaction volume that involves \textit{j}, or $p_{ij}=\frac{V_{ij}}{\sum_{j=1}^{k}V_{ij}}$, where $V_{ij}$ is the interaction volume between node \textit{i} and \textit{j}. Then, social diversity is defined as Shannon's entropy associated with individual \textit{i}’s communication behavior, normalized by \textit{k}: $D_{social}(i)=\frac{-\sum_{j=1}^{k}p_{ij}\log(p_{ij})}{\log(k)}$~\cite{eagle2010network}. 
 
\paragraph{Grouping Metrics} 

The correlations between properties of adjacent network nodes are known in the ecology and epidemiology literature as ``assortative mixing". If a node tends to establish edges with nodes that present some similarity with it (scoring groups in our setting), then they are said to present an assortative mixing, otherwise is it a disassortative mixing. In our directed network some rules are satisfied: $\sum_{ij} e_{ij}=1; \sum_{j} e_{ij}=a_{i}; \sum_{i} e_{ij}=b_{j}$, where $e_{ij}$ is the fraction of edges that connect a vertex of type $i$ to another of type $j$ and $a_{i}$ and $b_{i}$ are the fraction of each type of end of an edge attached to vertices of type $i$. The assortativity coefficient is defined as: $r=\frac{\sum_{i} e_{ii}-\sum_{i} a_{ij}b_{ij}}{1-\sum_{i} a_{ij}b_{ij}}$ (see \cite{newman2003mixing} for more details).

\paragraph{Principal Component Analysis}

After subtracting the mean value for each dimension, we calculated the covariance matrix for three dimensions: number of connections, interactions and the calculated topological diversity. We found that the eigenvalue proportions were $\simeq 0.3$ for all three eigenvalues. 

%\paragraph{Student Classification}

%Students were classified into low performing (score is one standard deviation below the mean), high performing (score one standard deviation above the mean) and mid. Thus, students were grouped in high (score $> 6.5$), low (score $< 3.5$) and mid (scores in between $3.5$ and $6.5$). 

\paragraph{Interaction Classification}

%Iterations themselves where classified: 1) those that render a sustained communication pattern ({\em persistent} or collaborative interactions) and 2) those that can be classified as exploratory in which a single response or no response at all was received ({\em transient} interactions).

\begin{figure}[htp]
\centering
\caption{(Top) Average number of total persistent interactions per group, (Bottom) Average number of transient interactions per scoring group. $^{*}  p<0.05$ as compared to low;$^{\#} p<0.05$ as compared to mid; n=290.}
\includegraphics[width=0.5\textwidth]{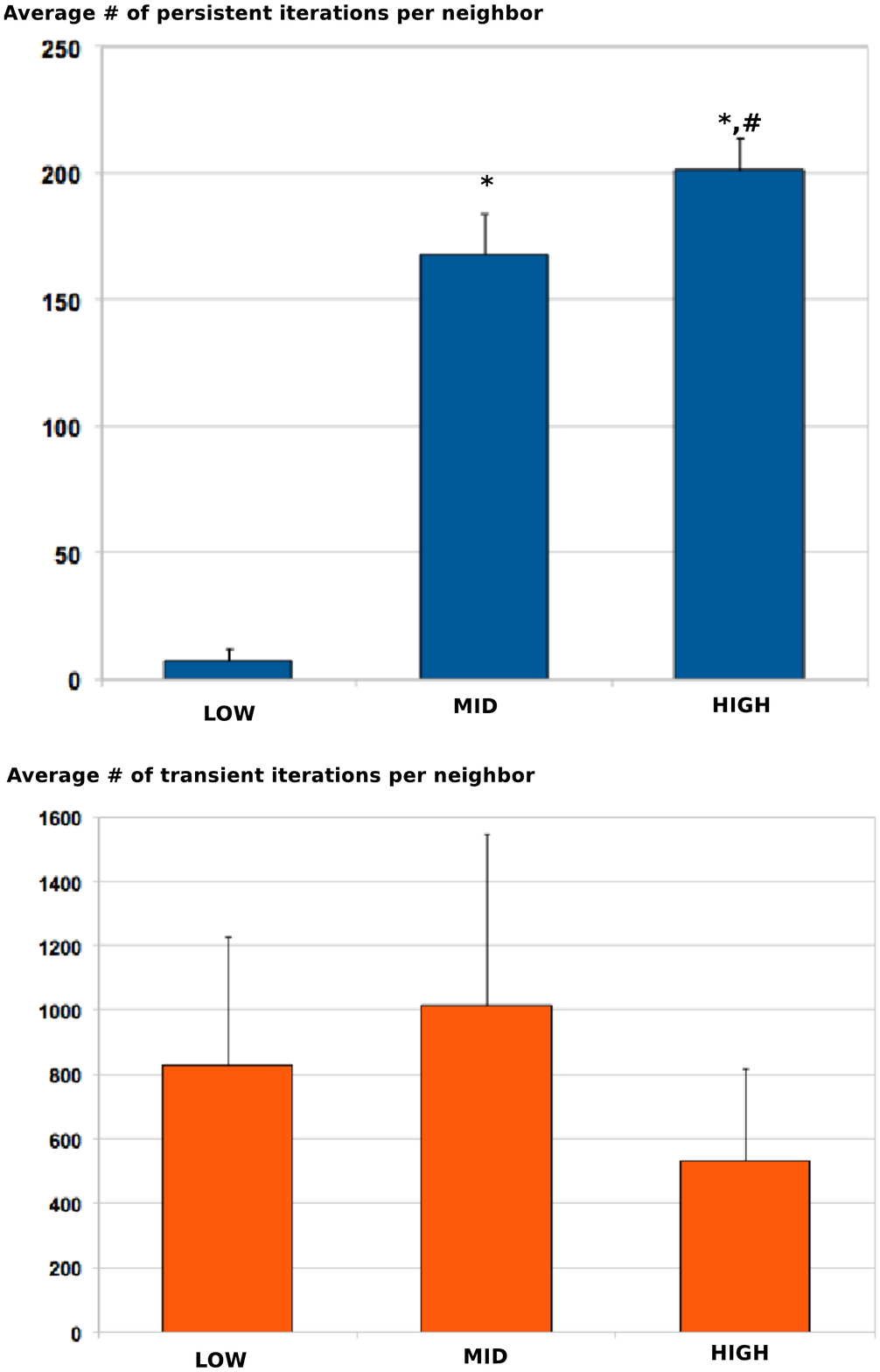}
\label{fig:data2}
\end{figure}

An interaction between students $i and j$ was classified as persistent if the contact ouccured at least twice (see Figure \ref{fig:data2}).

\paragraph{Temporal Analysis}

We first normalized the number of total interactions of the three courses to their respective maximum value of all weeks, in order to obtain a representative trace of the temporal course of the appearance of interactions. Then, we plotted the percentage of persistent interactions for all three groups of students as a function of time. We fitted these curves with a sigmoid function $ y = \frac{a}{(1.0 + exp(-\frac{(x-b)}{c}))}$, where $c$ represents the slope of change and $b$ the midpoint where approximately 50\% of the maximum value is reached. See the obtained results in Table \ref{table:temp}. For error minimization we set $a, b$ and $c$  to minimizr the lowest sum of squared absolute error, resulting in a value that was consistently lower than 0.026.

\begin{center}
\begin{table*}[t!]
\centering
\begin{tabular}{|l|l|l|l|l|l|}
\hline
 & \textbf{LOW} & \textbf{MID} & \textbf{HIGH} & \textbf{Total}\\ \hline
\textit{a} & 0.9 & 13.6 & 26.79 & 0.973\\ \hline
\textit{b} & 6.08 & 4.81 & 3.2 & 5.007\\ \hline
\textit{c} & 1.43 & 0.44 & 0.58 & 0.4769 \\  \hline
\end{tabular}
\caption{Sigmoid Fitting Results.}
\label{table:temp}
\end{table*}
\end{center}

\paragraph{Information Cascades}

Students were highly encouraged to use a systematic naming mechanisms for their files (First author name, year, 4-5 first words of the title). We employed data from BSCW and an automated mechanism (bash script) to recursively determine the files in the HOME directory for each user. 

From this information we created a subgraph, where students are the nodes and links represent the ``transfer'' of a file from a user's account to another's. A directed student to student edge is weighted with the total number of links
occurring between documents in source student and documents in the HOME of the destination student. Associated with each document is also the time, so we labelled the edges with the time difference $\delta$ between the appearance of the document in the HOME of the source and the destination. Let $t_{u}$ and $t_{v}$ denote appearance times of a document in the HOME of students $u$ and $v$, then $\delta  = t_{u} - t_{v}$, where $\delta > 0$ since there are no self-edges.

These subgraphs lead to information cascades, which are induced subgraphs by edges representing the ﬂow of information. We assumed this flow of information depends on the existence of an edge in the interaction graph. In other words, those students have interacted within the previous 72h\footnote{Please note that the assignments in this course typically consisted on weekly works, so a smaller time adds little information, while longer periods result in too many events to establish reasonable causality relationships.} prior to the appearance of the document in the HOME of the destination student.

% Do NOT remove this, even if you are not including acknowledgments
\subsection*{Acknowledgments}

We would like to thank Charles Elkan and Miranda Mowbray for their insightful comments on the manuscript.

%\section*{References}
% The bibtex filename
\bibliography{plos2012}

%\subsection*{Figure Legends}
%\begin{figure}[!ht]
%\begin{center}
%%\includegraphics[width=4in]{figure_name.2.eps}
%\end{center}
%\caption{
%{\bf Bold the first sentence.}  Rest of figure 2  caption.  Caption 
%should be left justified, as specified by the options to the caption 
%package.
%}
%\label{Figure_label}
%\end{figure}

%\subsection*{Tables}
%\begin{table}[!ht]
%\caption{
%\bf{Table title}}
%\begin{tabular}{|c|c|c|}
%table information
%\end{tabular}
%\begin{flushleft}Table caption
%\end{flushleft}
%\label{tab:label}
% \end{table}

\end{document}